\documentclass{ws-procs961x669}            % default, citations in superscript

\usepackage{physics}
\usepackage{hyperref}

\begin{document}
\title{Polymer Quantization of the Isotropic Universe:\\comparison with the Bounce of Loop Quantum Cosmology}

\author{G. Barca$^*$, E. Giovannetti and F. Mandini}

\address{Department of Physics, La Sapienza University of Rome, Rome, Italy\\
$^*$Speaker. E-mail: gabriele.barca@uniroma1.it}

\author{G. Montani}

\address{Department of Physics, La Sapienza University of Rome, Rome, 00185 Italy\\
Fusion and Nuclear Safety Department, ENEA, Frascati (RM), Italy}

\begin{abstract}
We implement Polymer Quantum Mechanics on the Hamiltonian formulation of the isotropic Universe in both the representations of the standard Ashtekar-Barbero-Immirzi connection and of a new generalized coordinate conjugate to the Universe volume. The resulting dynamics is a bouncing cosmology; when quantizing the volume-like variable the Big Bounce is an intrinsic cut-off on the cosmological dynamics, while when using the standard connection the Bounce density results to be dependent on the initial conditions of the prepared wave packet. Then we compare the nature of the resulting Bounce with what emerges in Loop Quantum Cosmology, where the dependence of the critical density on the initial conditions is present when the minimum area eigenvalue is implemented in a comoving representation instead of the physical one. We conclude that, if one hand the preferable scenario should be a Big Bounce whose density depends on initial conditions in view of the privileged SU(2) character that the Ashtekar-Barbero-Immirzi connection possesses in the full Loop Quantum Gravity, on the other hand the equivalence demonstrated in the context of polymer cosmology can be a hint in favour of the viability of the improved scheme of Loop Quantum Cosmology even though it is not expressed through the privileged set of variables.
\end{abstract}


\bodymatter

\section{Introduction}
Loop Quantum Cosmology (LQC)\cite{BojowaldOriginalLQC,Ashtekar0,Ashtekar1,Ashtekar2,Ashtekar3} is the implementation of Loop Quantum Gravity (LQG)\cite{ROVELLI199080} in the symmetry-reduced cosmological minisuperspaces. It predicts the emergence of a Bouncing scenario that removes the classical singularity, but has the intrinsic limitation that the $SU(2)$ symmetry of LQG is lost.

Here we address the nature of the Bounce through the implementation of Polymer Quantum Mechanics (PQM)\cite{CorichiPQM} on the flat isotropic Friedmann-Lema\^itre-Robertson-Walker (FLRW) model in two different representations: the Ashtekar-Barbero-Immirzi variables (the basic $SU(2)$ variables of LQG and LQC) and the volume variables\cite{Mantero}. Then, we compare our results with LQC in order to gain some insight on both frameworks. A more detailed analysis is presented in Ref.~\citenum{Mandini}.

This work is organized as follows: in section \ref{secPQM} we introduce the framework of PQM; then we implement PQM on the FLRW model firstly expressed in the Ashtekar variables in section \ref{Ashtvar}, and then in the volume variables in section \ref{volvar}; in section \ref{canonequiv} we compare our results with LQC; finally in section \ref{concl} we summarize our work. We use the natural units $\hslash=c=8\pi G=1$.

\section{Polymer Quantum Mechanics}
\label{secPQM}
PQM\cite{CorichiPQM} is an alternative representation that is non-unitarily connected to the standard Schr\"odinger representation. It implements a fundamental scale in the Hilbert space through the introduction of a lattice structure. Its aim is to reproduce effects similar to those of LQG through an independent framework that allows for more freedom in the choice of the preferred configurational variable and that is easily and reliably applicable to many Hamiltonian systems.

Let us consider a generic Hamiltonian system with canonical variables $(Q,P)$. The kinematical framework of PQM consists in assigning to one of the variables, usually the position $Q$, a discrete character. As a consequence, in the quantization procedure, the conjugate momentum $P$ cannot be promoted to a well-defined operator because the translational operator $\hat{T}(\lambda)=e^{i\lambda P}$ is not weakly continuous.

The procedure to regularize the momentum $P$ and to construct a well-defined operator consists in the introduction of a lattice on position $Q$ with constant spacing $\beta_0\in\mathbb{R}$. The Hilbert space is then constructed as the one that contains all those states $\ket{\psi}=\sum_i\,b_i\,\ket{\beta_n}$, with $\beta_n=n\beta_0$ and $\sum_i\abs{b_i}^2<\infty$. Now the translational operator must be restricted to act only by discrete steps to remain on $\gamma_{\beta_0}$: this is achieved simply by setting $\zeta=\beta_0$ in order to obtain $\hat{T}(\beta_0)\,\ket{\beta_n}=\ket{\beta_{n+1}}$.

When the condition $P\beta_0\ll1$ is satisfied, we can approximate $P$ with a sine function, so that the regulated momentum operator can be constructed as the incremental ratio of the translational operator on the lattice:
\begin{subequations}
\begin{equation}
    P\approx\frac{\sin(\beta_0 P)}{\beta_0}=\frac{e^{i\beta_0P}-e^{-i\beta_0P}}{2i\beta_0},
    \label{ppoly}
\end{equation}
\begin{equation}
    \widehat{P_{\beta_0}}\ket{\beta_n}=\frac{\hat{T}(\beta_0)-\hat{T}(-\beta_0)}{2i\beta_0}\,\ket{\beta_n}=\frac{\ket{\beta_{n+1}}-\ket{\beta_{n-1}}}{2i\beta_0}.
    \label{hatppoly}
\end{equation}
\end{subequations}
For the squared momentum operator there are infinitely many approximations, but they usually amount to just a rescaling of the lattice parameter $\beta_0$; the simplest one in order to avoid discrepancies is obtained by composing the operator \eqref{hatppoly} with itself, corresponding to the approximation $P^2\approx\sin^2{(\beta_0 P)}/\beta^2_0$.

When performing the quantization of a system using the momentum polarization of the polymer representation, the regulated momentum operator must be used together with the differential coordinate operator. Alternatively, it is possible to perform a semiclassical analysis by using the formal substitution \eqref{ppoly} in the classical Hamiltonian, thus including quantum modifications in the classical dynamics \cite{MoriconiAniso2017,Antonini,Mantero}.

\section{Polymer FLRW Dynamics in Ashtekar Variables}
\label{Ashtvar}
We will now show the dynamics of the flat isotropic FLRW model expressed in the Ashtekar variables. Given the symmetries of the model, the connection and the triads reduce to a very simple form: $A^i_a\to c=\gamma\dot{a}$, $E^a_i\to p$, with $\abs{p}=a^2$, $\pb{c}{p}=\gamma/3$, where $a=a(t)$ is the scale factor of the Universe, the dot represents a derivative with respect to time $t$ and $\gamma$ is the Immirzi parameter.

\subsection{Semiclassical Dynamics}
We consider the FLRW model filled with a free massless scalar field $\phi$; we discretize the area $p$, so we use the sine substitution \eqref{ppoly} on the connection $c$. We have a four-dimensional phase space $(c,p;\phi,P_\phi)$, where $P_\phi$ is the momentum conjugate to the scalar field. The modified Hamiltonian constraint for this model is
\begin{equation}
    \mathcal{C}_\text{poly}=-\frac{3}{\gamma^2\beta_0^2}\,\sqrt{p}\,\sin[2](\beta_0c)+\rho_\phi\,p^\frac{3}{2}=0.
\end{equation}
Then the dynamics is derived through the standard Hamilton's equations.

Through the constraint, we obtain a modified Friedmann equation:
\begin{equation}
    H^2=\left(\frac{\dot{a}}{a}\right)^2=\left(\frac{\dot{p}}{2p}\right)^2=\frac{\rho_\phi}{3}\left(1-\frac{\rho_\phi}{\rho_\beta}\right),\quad\rho_\beta=\frac{3}{\gamma^2\beta_0^2p};
\end{equation}
the correction factor in parentheses contains a regularizing density $\rho_\beta$ that depends on time through $p$; the minus sign introduces a critical point for the dynamics.

Since $P_\phi$ is a constant of motion, it is possible to use the scalar field as internal time; this makes the Friedmann equation easily solvable, and its solution $p(\phi)$ is
\begin{equation}
    p(\phi)=\frac{\gamma\beta_0P_\phi}{\sqrt{6}\,}\,\cosh\left(\sqrt{\frac{2}{3}\,}\,\phi\right).
\end{equation}
The hyperbolic cosine has a non-zero minimum: the singularity is avoided and the Big Bang is replaced by a Big Bounce. However, the critical density, i.e. the energy density at the Bounce, is $\rho_\text{crit}=\rho_\beta(p|_{\phi_B})=3^{\frac{3}{2}}\,\sqrt{2\,}\,/P_\phi\gamma\beta_0\,$: it is dependent on the constant of motion $P_\phi$ that must be therefore set through initial conditions.

\subsection{Quantum Dynamics}
\begin{figure}
    \centering
    \includegraphics[scale=0.48]{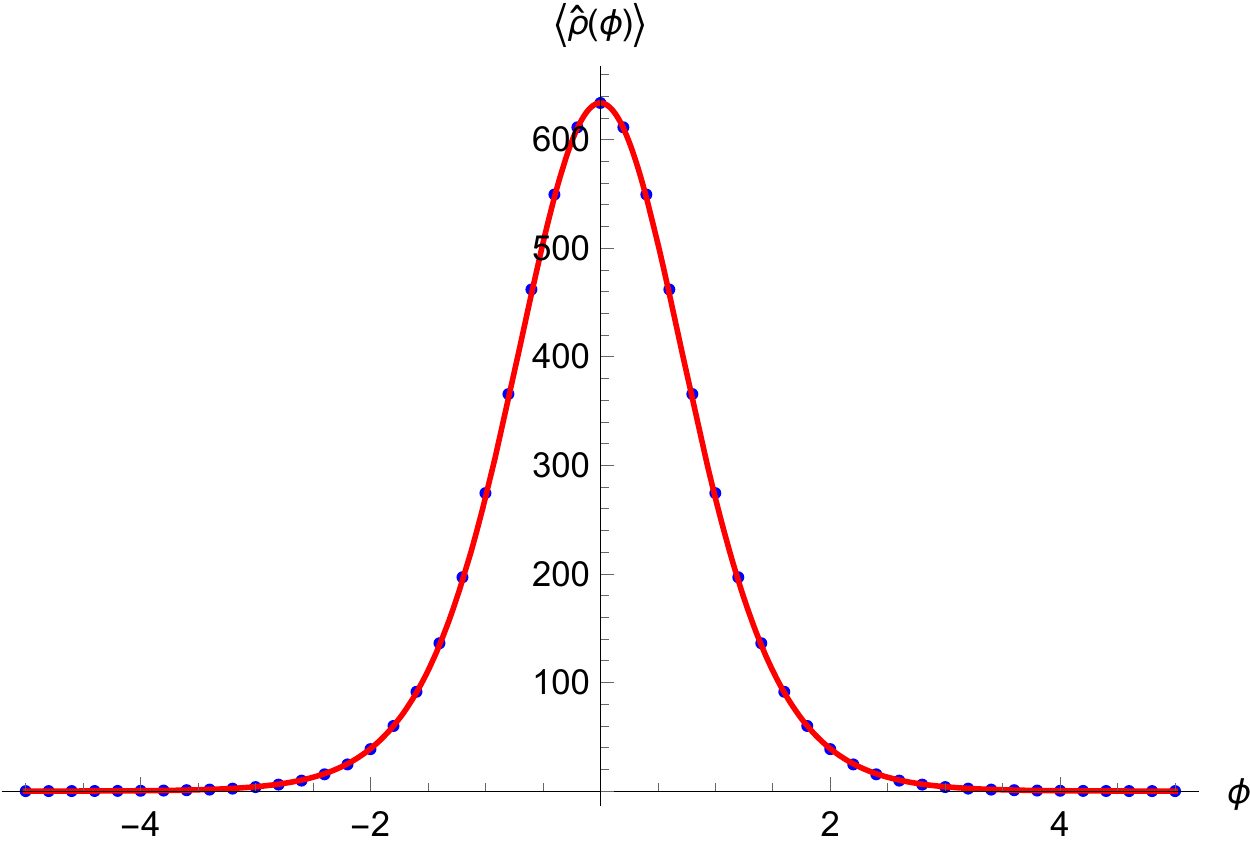}
    \,
    \includegraphics[scale=0.48]{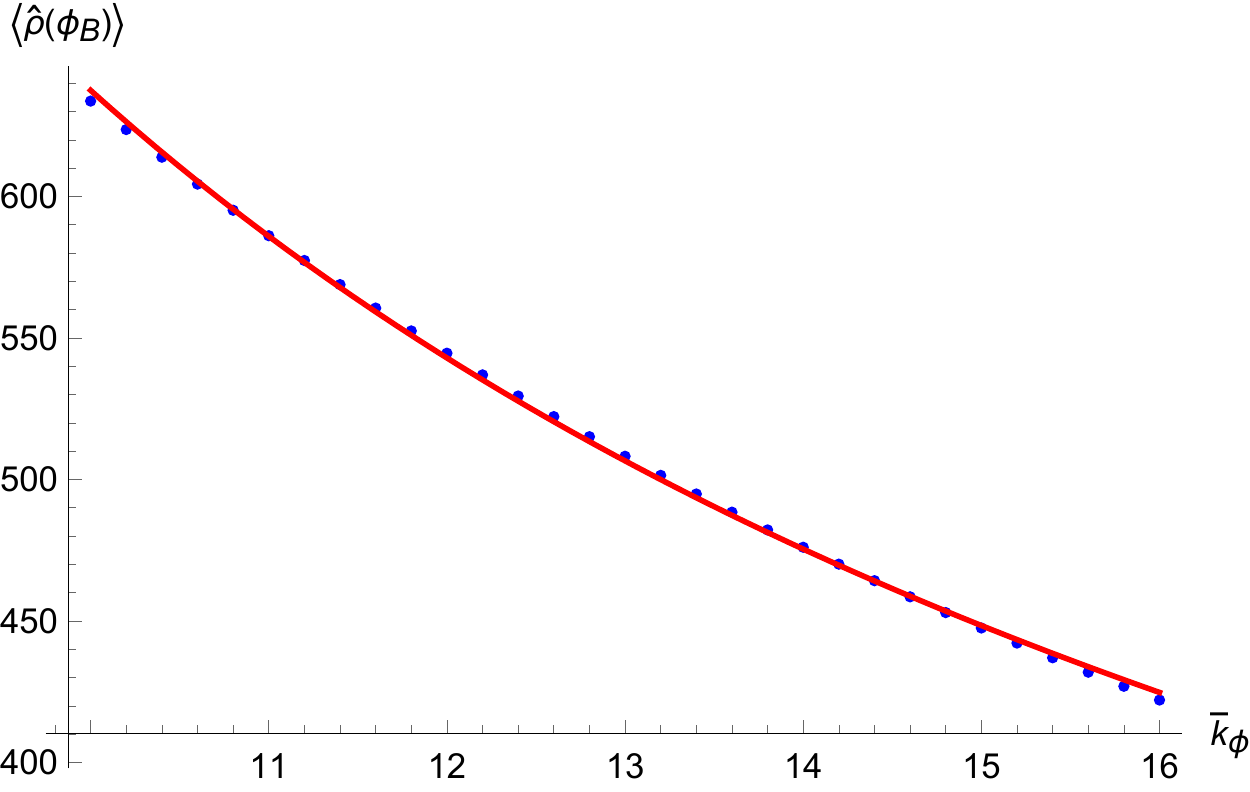}
    \caption{Behaviour of the expected value of the energy density operator $\ev{\widehat{\rho_\phi}}$ as function of time $\phi$ (left panel) and dependence of the value of the energy density at the Bounce $\ev{\widehat{\rho_\phi}}\big|_{\phi_B}$ on the initial condition for the wavepacket $\overline{k}§_\phi$ (right panel). The blue dots are the quantum expectation values, fitted with red continuous functions in accordance with semiclassical evolution.}
    \label{rhopcquantum}
\end{figure}

In the quantum picture, the fundamental variables are promoted to operators so that the quantum Hamiltonian constraint yields a Wheeler--De Witt (WDW) equation:
\begin{subequations}
\begin{equation}
    \hat{c}=\frac{\sin(\beta_0c)}{\beta_0},\quad\hat{p}=-i\frac{\gamma}{3}\dv{c},\quad\widehat{P_\phi}=-i\dv{\phi},
\end{equation}
\begin{equation}
    \hat{\mathcal{C}}_\text{poly}\Psi(c,\phi)=\left[-\frac{2}{3\beta_0^2}\left(\sin(\beta_0c)\,\dv{c}\right)^2+\dv[2]{\phi}\right]\Psi(c,\phi)=0;
    \label{Cpolyquantum}
\end{equation}
\end{subequations}
through a substitution, the latter is recast as a Klein-Gordon-like (KG) equation:
\begin{equation}
    x=\sqrt{\frac{3\,}{2\,}\,}\,\ln\left[\tan\left(\frac{\beta_0c}{2}\right)\right]+x_0\quad\implies\quad\dv[2]{x}\Psi(x,\phi)=\dv[2]{\phi}\Psi(x,\phi).
\end{equation}
We construct its solution as a Gaussian-like wavepacket over eigenvalues $k_\phi$ of $\hat{P}_\phi$:
\begin{equation}
    \Psi(x,\phi)=\int_0^\infty\frac{\dd k_\phi}{\sqrt{4\pi\sigma}}\,e^{-\frac{(k_\phi-\overline{k}_\phi)^2}{2\sigma^2}}\,k_\phi\,e^{ik_\phi x}\,e^{ik_\phi\phi};
\end{equation}
then we evolve it backwards in time, and study the behaviour of the expectation value of the energy density operator through a KG scalar product of the form
\begin{equation}
    \ev{\widehat{\rho_\phi}}=\int_{-\infty}^\infty\dd x\,\Big(\Psi^*\partial_\phi(\widehat{\rho_\phi}\Psi)-(\widehat{\rho_\phi}\Psi)\partial_\phi\Psi^*\Big),\quad\widehat{\rho_\phi}=\frac{\hat{P}_\phi^2}{2\abs{\hat{p}}^3}.
    \label{KGproduct}
\end{equation}
The result is a function of time $\phi$. Figure \ref{rhopcquantum} shows the energy density increasing towards the singularity, reaching a finite maximum and then decreasing again; in view of its scalar nature, this behaviour confirms the presence of a Big Bounce also at the quantum level. However, the critical density at the Bounce (i.e. the value of the peak) is dependent on the initial condition $\overline{k}_\phi$, as in the semiclassical dynamics.

\section{Polymer FLRW Dynamics in Volume Variables}
\label{volvar}

\subsection{Specialization of Polymer Cosmology}
In order to find the suitable variable to obtain a fixed Bounce density\cite{Mantero}, we perform a canonical transformation on the classical Hamiltonian to a generic function $F$ of the scale factor: $p\to F(a)$, $c\to P_F=2\sqrt{p\,}\,c\,/F'(a)$ with $F'(a)=\partial F/\partial a$. By using the polymer substitution \eqref{ppoly} on the new momentum $P_F$, we obtain a modified Hamiltonian constraint and a corresponding modified Friedmann equation:
\begin{subequations}
\begin{equation}
    \mathcal{C}_\text{poly}=-\frac{3}{4\gamma^2\beta_0^2}\,\frac{\Big(F'(a)\Big)^2}{\sqrt{p\,}}\sin[2](\beta_0P_F)+\rho_\phi\,p^\frac{3}{2}=0,\quad\rho_\phi=\frac{P_\phi^2}{2p^3},
\end{equation}
\begin{equation}
    H^2=\frac{\rho_\phi}{3}\left(1-\frac{\rho_\phi}{\tilde{\rho}_\beta}\right),\quad\tilde{\rho}_\beta=\frac{3}{4\gamma^2\beta_0^2}\,\frac{\Big(F'(a)\Big)^2}{p^2}\propto\left(\frac{F'(a)}{p}\right)^2.
\end{equation}
\end{subequations}
To obtain a universal regularizing density $\tilde{\rho}_\beta$, we must have $F(a)\propto a^3$; we choose to use as new fundamental variables $v=a^3$ and $\tilde{c}=2c/3\sqrt{p\,}\propto\dot{a}/a$.

\subsection{Semiclassical Dynamics}
We implement PQM on the system by considering $v$ as discrete and using the formal substitution \eqref{ppoly} on the new generalized coordinate $\tilde{c}$:
\begin{equation}
    \mathcal{C}_\text{poly}=-\frac{27}{4\gamma^2\beta_0^2}\,v\,\sin[2](\beta_0\tilde{c})+\rho_\phi\,v=0,\quad\rho_\phi=\frac{P_\phi^2}{2v^2}.
\end{equation}
This new representation yields a modified Friedmann equation with a fixed regularizing density, as expected:
\begin{equation}
    H^2=\left(\frac{\dot{a}}{a}\right)^2=\left(\frac{\dot{v}}{3v}\right)^2=\frac{\rho_\phi}{3}\left(1-\frac{\rho_\phi}{\tilde{\rho}_\beta}\right),\quad\tilde{\rho}_\beta=\frac{27}{4\gamma^2\beta_0^2}=\tilde{\rho}_\text{crit}.
\end{equation}
The quantity $\tilde{\rho}_\beta=\tilde{\rho}_\text{crit}$ depends only on fundamental constants and the Immirzi parameter, and can therefore be already considered a critical density. Using $\phi$ as time, the new modified Friedmann equation is easily solvable for $v(\phi)$:
\begin{equation}
    v(\phi)=\frac{2\gamma\beta_0}{\sqrt{54\,}}\,P_\phi\cosh\left(\sqrt{\frac{3}{2}\,}\,\phi\right).
\end{equation}
We still have a hyperbolic cosine, indicating a  Bounce that replaces the Big Bang, but this time the value of the energy density at which this happens is universal.

\subsection{Quantum Dynamics}
In the quantization procedure the fundamental variables are promoted to operators and the quantum Hamiltonian constraint yields a WDW equation:
\begin{equation}
    \hat{v}=-i\,\dv{\tilde{c}},\quad\hat{\tilde{c}}=\frac{\sin(\beta_0\tilde{c})}{\beta_0},\quad\hat{P}_\phi=-i\,\dv{\phi};
\end{equation}
\begin{equation}
    \hat{\tilde{\mathcal{C}}}_\text{poly}\Psi(\tilde{c},\phi)=\left[-\frac{3}{2\beta_0^2}\left(\sin(\beta_0\tilde{c})\,\dv{\tilde{c}}\right)^2+\dv[2]{\phi}\right]\Psi(\tilde{c},\phi)=0.
\end{equation}
Through a suitable substitution, the latter is recast as a KG-like equation in the variable $\tilde{x}$ where $\phi$ plays the role of time:
\begin{equation}
    \tilde{x}=\sqrt{\frac{2\,}{3\,}\,}\,\ln\left[\tan\left(\frac{\beta_0\tilde{c}}{2}\right)\right]+\tilde{x}_0\quad\implies\quad\dv[2]{\tilde{x}}\Psi(\tilde{x},\phi)=\dv[2]{\phi}\Psi(\tilde{x},\phi).
\end{equation}
We construct the solution as a Gaussian-like wavepacket of the form
\begin{equation}
    \Psi(\tilde{x},\phi)=\int_0^\infty\frac{\dd k_\phi}{\sqrt{4\pi\sigma}}\,e^{-\frac{(k_\phi-\overline{k}_\phi)^2}{2\sigma^2}}\,k_\phi\,e^{ik_\phi\tilde{x}}\,e^{ik_\phi\phi}.
\end{equation}
Finally, we evolve it backwards in time and study the behaviour of the expected value of relevant operators (the volume $\hat{V}=\hat{v}$ and the energy density $\widehat{\rho_\phi}=\hat{P}_\phi^2/2\hat{v}^2$) through the same KG scalar product \eqref{KGproduct} in terms of $\tilde{x}$.
\begin{figure}
    \centering
    \includegraphics[scale=0.48]{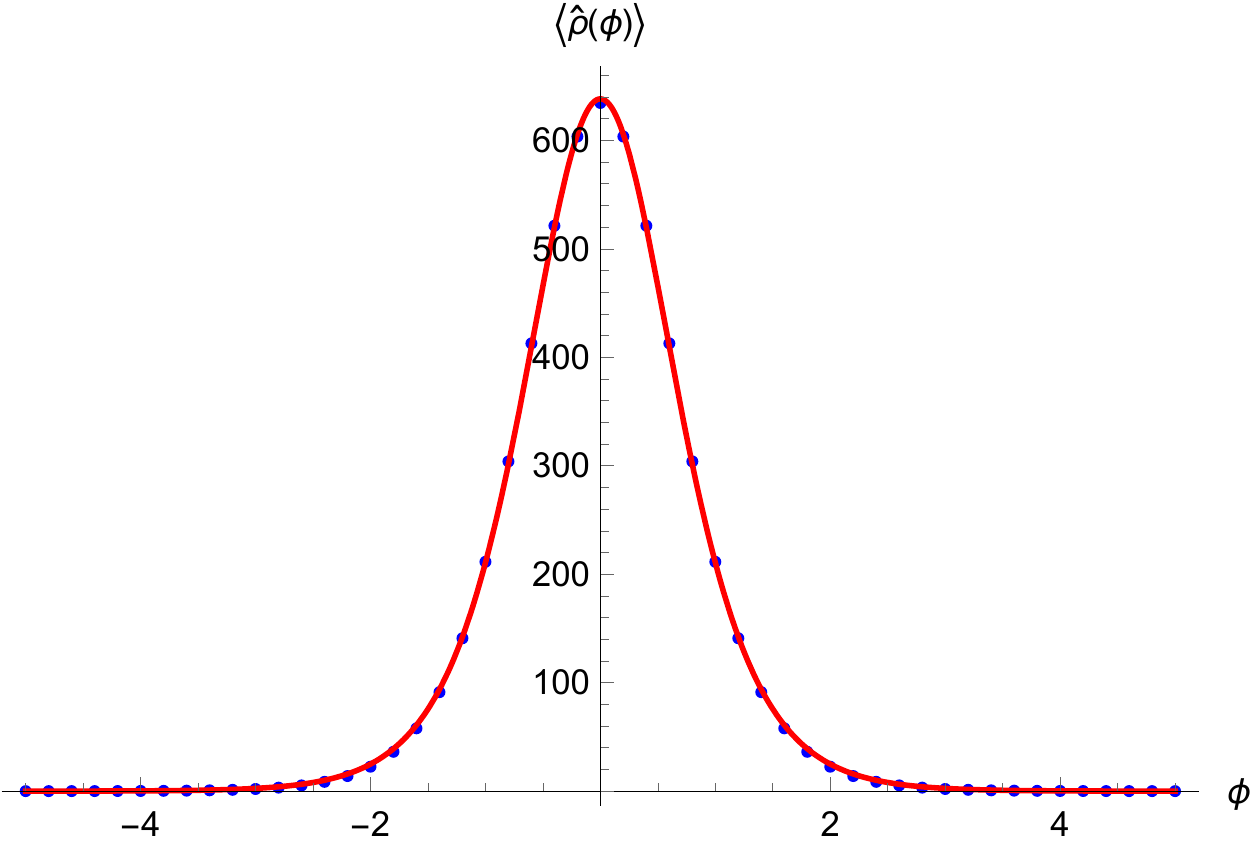}
    \,
    \includegraphics[scale=0.48]{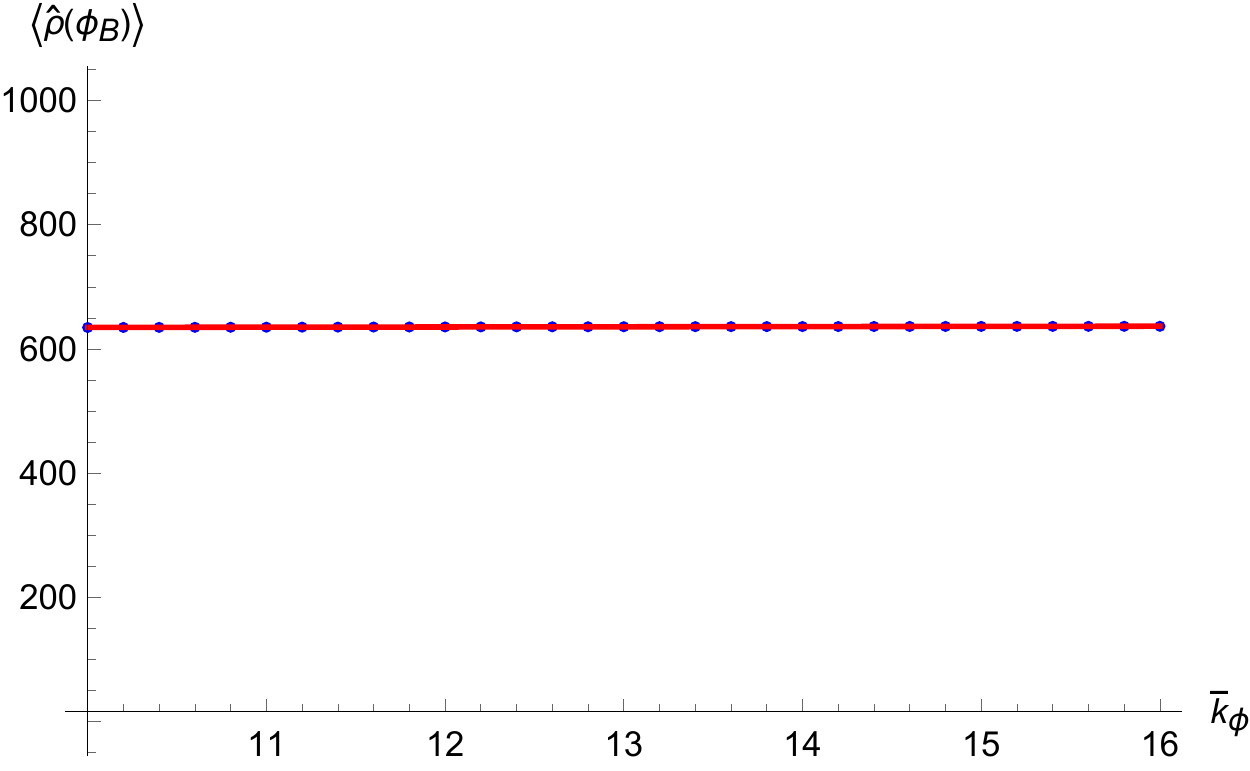}
    \caption{Expected value of the energy density operator as function of time $\phi$ (left panel) and dependence of the critical density (the value of the energy density at the Bounce) on the initial condition for the wavepacket $\overline{k}_\phi$. The blue dots are the quantum expectation values that have been fitted with red continuous functions in accordance with semiclassical evolution.}
    \label{rhovc}
\end{figure}
The Bounce is present also in this quantum setting; however in this case the energy density, depicted in the left panel of figure \ref{rhovc}, has a value at the Bounce that is constant and independent on $\overline{k}_\phi$, as shown in the right panel of the same figure.

\section{Discussion of the Results}
\label{canonequiv}

\subsection{The Nature of the Bounce in Polymer Cosmology}
The quantization procedure of PQM implements on a Hamiltonian system a lattice that has a constant step by construction. Therefore the canonical changes of variables must be performed on the classical Hamiltonian constraint before discretization. The nature of the resulting Bounce in the two pictures is not the same, and the two dynamics are inequivalent.

To recover the equivalence, we can perform the canonical transformation \emph{after} the implementation of the lattice. We consider the system with the discretized volume and change the variables to Ashtekar ones; the canonical transformation must preserve the Poisson brackets and therefore must satisfy the condition $\beta_0\tilde{c}=\beta'c$. We obtain a discretized system in the variables $(c,p)$ that has a new lattice spacing $\beta'$ dependent on the variable $p$. The new equation of motion for $p$ is mapped to the original equation of motion in the Ashtekar variables, but the lattice parameter $\beta_0$ is replaced by the new one $\beta'$; the resulting dynamics is a Big Bounce scenario with a fixed, universal critical density, the same obtained by discretizing the volume with a constant lattice. Thus we conclude that the nature of the Bounce is decided by the variable to which a constant lattice spacing is assigned.

We stress that this result can be demonstrated only at a semiclassical level; on the quantum level this comparison is impossible, because a quantization procedure with a non-constant translational parameter has not been developed yet.

\subsection{Comparison with Loop Quantum Cosmology}
One of the most important results of LQG is the quantization of the kinematical geometrical operators of area and volume\cite{RovelliAreaVolume}, with the Area Gap $\Delta$ being the smallest non-zero area eigenvalue. In the construction of the quantum Hamiltonian constraint of LQC, the area gap is introduced from the full theory somewhat ad hoc, and the character of its implementation defines different schemes of LQC. In the first formulation, the area gap was implemented as a kinematical feature (as in full LQG) on the comoving area: this is equivalent to introducing a fixed lattice and yields the original $\mu_0$ scheme of LQC, a scenario very similar to our polymer cosmology with the Ashtekar variables of section \ref{Ashtvar}, that presents a Big Bounce dependent on initial conditions\cite{Ashtekar0,Ashtekar1}. Then, in the improved dynamics, the area gap is implemented as a dynamical feature on the physical area, i.e. rescaled by the squared scale factor; this leads to a lattice with a non-constant spacing $\bar{\mu}(p)$, and prevents quantization. Therefore a change of basis is performed to a volume-like variable $\nu\propto v$ to make the lattice parameter constant (note how this is the inverse change of variables that we performed above). Now the quantization can be completed, and the resulting dynamics is a Big Bounce with a critical density that is a universal feature \cite{Ashtekar2,Ashtekar3}. This scenario, dubbed $\bar{\mu}$ scheme, is analogous to the polymer FLRW cosmology in volume variables presented here in section \ref{volvar}.

The improved scheme avoids the dependence of quantum gravitational effects on initial conditions, a feature that does not have a very clear physical interpretation, and therefore seems to be more appealing than the original formulation. One criticism is that the new $\bar{\mu}$ scenario takes LQC further away from LQG because it does not use the Ashtekar variables anymore, that are privileged in the full theory in view of their $SU(2)$ character. However the dynamical similarities of PQM and LQC may justify the change of variables: at least on a semiclassical effective level, the dynamics of the model expressed in volume variables with a constant spacing is equivalent to that of the same model expressed in Ashtekar variables with a $p$-dependent lattice step.

\section{Summary and Conclusions}
\label{concl}
We have shown how the implementation of PQM on the flat isotropic model solves the singularity by replacing the Big Bang with a Big Bounce, but its nature depends on the fundamental variable used to describe the model: with the area $p$ from the Ashtekar variables the energy density at the Bounce is dependent on the initial conditions, while if the volume $v=a^3$ is used instead, the Bounce is a universal feature with a fixed energy density; therefore the two pictures yield two different and inequivalent dynamics. A possibility to recover the equivalence is to perform the canonical transformation after the implementation of the lattice and to allow for a non-constant step; this does not affect the dynamics, meaning that the nature of the Bounce is decided by the variable that has a lattice with constant spacing.

In LQC two analogous pictures emerge: the original $\mu_0$ scheme yields a Bounce dependent on initial conditions, while the improved $\bar{\mu}$ scheme prevents this feature through a change of basis and presents a universal Bounce. These similarities with polymer cosmology may be used to gain insight into the validity of the two scenarios: if on one hand the improved scheme, although physically more appealing, can be criticized for not using the privileged $SU(2)$ variables of LQG, on the other hand the parallelism with the minisuperspace implementation of PQM justifies the $\bar{\mu}$ scenario in view of the equivalence demonstrated in semiclassical polymer cosmology.

\bibliographystyle{ws-procs961x669}
\bibliography{bib}

\end{document}